\documentclass{ws-ijmpd}
\usepackage[super,compress]{cite}
\usepackage{setspace, longtable, amstext}
\begin{document}

\markboth{Authors' Names}
{Instructions for Typing Manuscripts (Paper's Title)}

%
\catchline{}{}{}{}{}
%

\title{Satellites testing general relativity: Residuals versus perturbations}  

\author{V. G. Gurzadyan$^1$\footnote{gurzadyan@yerphi.am}, I. Ciufolini$^2$, A.Paolozzi$^3$,
A.L. Kashin$^1$, H.G. Khachatryan$^1$, S. Mirzoyan$^1$ and G. Sindoni$^3$}

\address{1. Center for Cosmology and Astrophysics, Alikhanian National Laboratory and Yerevan State University, Yerevan, Armenia\\
2.Dipartimento di Ingegneria dell'Innovazione, Universit\`a del Salento, Lecce and Museo Storico della Fisica e
Centro Studi e Ricerche Enrico Fermi, Rome, Italy\\
3. Scuola di Ingegneria Aerospaziale, Sapienza Universit\`a di Roma and Museo Storico della Fisica e
Centro Studi e Ricerche Enrico Fermi, Rome, Italy}

\maketitle

\begin{history}
\received{15 December 2016}
\accepted{15 January 2017}
\end{history}

\begin{abstract}
Laser ranging satellites have proved their efficiency for high precision testing of the effect of frame-dragging, one of remarkable predictions of the General Relativity. The analysis of the randomness properties of the residuals of LAGEOS and LAGEOS 2 satellites reveals the role of the
thermal thrust -- Yarkovsky effect -- on the satellite which was in the orbit for longer period (LAGEOS). We also compute Earth's tidal modes affecting the satellite 
LARES. The recently obtained 5\% accuracy limit reached for the frame dragging effect based on the 3.5 year data of LARES analysed together with those of LAGEOS satellites and using the Earth gravity model of GRACE satellite, is also represented. 
\end{abstract}


\ccode{PACS numbers: 04.20.-q}


\section{Introduction}	

General Relativity (GR) since its creation a century ago has undergone a number of remarkable experimental tests \cite{CW}.  At the same time, the possibility for revealing the limits of application of GR remains of principal importance due to the links to fundamental physical concepts such as the equivalence principle and Lorentz invariance. Along with that, GR is the basis for interpretation of the observed accelerated expansion of the Universe, of its current structure and of the early evolution, of the processes ongoing in galactic nuclei involving black holes and compact stars. GR tests of ever increasing precision are used also for constraining various extensions of GR and, therefore, are among the priorities in ongoing experimental and observational/cosmological projects.

Laser ranging satellites - LAGEOS and LAGEOS 2 - have proved their efficiency in high precision probing of the frame-dragging effect, one of predictions of GR \cite{CP,C}. LARES, the recently launched higher-mean-density laser ranging satellite, provides the best ever created test particle to move via geodesics (Ehlers-Geroch theorem)\cite{M} and, hence, enables to increase further the accuracy of the measurement of the sought GR's effect \cite{L2012,L2013,L2016}.     

Refined treatment of perturbations of various nature influencing the laser ranging satellites is one of main tasks in the detection of the frame-dragging effect within the measured signals, and, hence, represents a broad and complex area of studies (see \cite{C}). The problem is
in the analysis of the perturbations vs the residuals between the measured and theoretically predicted orbits of the satellites.
Here, we will briefly concentrate on particular aspects regarding the perturbations, i.e. how the randomness properties of the residuals can reveal differences for signals for LAGEOS and LAGEOS 2 satellites similar with other characteristics. We consider the dominant modes for the Earth's tides influencing the laser ranging satellites. Then, we represent the outcome of the analysis of around 3.5-year
measurements for LARES combined with those of LAGEOS and LAGEOS 2 satellites, and using Earth's gravity field model produced by geodesy satellite GRACE \cite{L2016}.  

\section{Thermal thrust and the Kolmogorov function}

Yarkovsky effect predicted in early XXth century is associated to the additional thrust of the object due its one-sided heating by the Solar radiation. Since the thermoconductivity of laser ranging satellites is also finite, one expects that the thermal thrust (including the Rubincam-Yarkovsky effect due to heating from Earth's atmosphere) can also have its signature in the residuals of laser ranging satellites.  

Let us inquire into the statistical properties of the residuals of LAGEOS and LAGEOS 2 (Figure 1) using the concepts of stochasticity parameter and of degree of randomness as defined by Kolmogorov \cite{Kolm} and Arnold \cite{Arnold_MMS}.  The stochasticity parameter is defined for a sequence of real-valued variables and  
for theoretical $F(x)$ and empirical $F_n(x)$ distribution functions as  
\begin{equation}\label{KSP}
\lambda_n=\sqrt{n}\ \sup_x|F_n(x)-F(x)|\ .
\end{equation}

Then, the Kolmogorov function $\Phi(\lambda)$ is the limit for the probability 
\begin{equation}
\lim_{n\to\infty}P\{\lambda_n\le\lambda\}=\Phi(\lambda)\ ,
\end{equation}
and has the form 
\begin{equation}
\Phi(\lambda)=\sum_{k=-\infty}^{+\infty}\ (-1)^k\ e^{-2k^2\lambda^2}\ ,\ \  \lambda>0\ , \Phi(0)=0 \label{Phi}.
\end{equation}
Here, the limit is converged uniformly and the distribution $\Phi$ is independent on the function $F$.
 
The Kolmogorov function was estimated for the standard deviation of the residuals of the satellites, and from Figure 2 one can see that those of LAGEOS are more random as compared to those of LAGEOS 2. Such a difference in the behaviour of randomness of $\sigma$ is due to the different effect of thermal thrust on LAGEOS and LAGEOS 2. Indeed, the Yarkovsky thermal thrust assumes a constant spin axis direction of LAGEOS
and LAGEOS 2, and a fast enough (with respect to the orbital period) spin rate necessary for the latitudinal thermal gradient to be built
between the north and south hemispheres. However, the analysis of the LAGEOS and LAGEOS 2 orbits and of their residuals reported here was carried out between 1992 and 2004, during this period, the conditions of constant spin axis orientation and fast enough spin rate were satisfied by LAGEOS 2, that was launched in 1992, but not by LAGEOS launched in 1976. In other words, during the period of our analysis, the LAGEOS satellite had an extremely low spin rate and its spin orientation, far from being constant, was almost chaotic. In conclusion, LAGEOS satellite should not show any periodical effects in the nodal residuals due to thermal accelerations. On the other hand, LAGEOS 2 was spinning fast enough during that period of analysis and with constant spin orientation, so it should show periodical i.e. regular effects in the orbital residuals, as follows from Figure 2.\cite{EPL}  

The anisotropy of heating and, hence, of the relevant temperatures of the hemispheres lead to an additional acceleration directed along the spin axis of a spherical satellite and yields \cite{EPL}
\begin{equation}
a \, \simeq \, 4\pi r^2_S \varepsilon \frac{ \sigma  T^3 \, \Delta\, T}{m_S},
\end{equation} 
where $\varepsilon$ is the coefficient of emissivity of the satellite of radius $r_S$ and mass $m_S$, $\sigma$ is Stefan's constant, and $\Delta T$ is the difference of temperatures of the hemispheres. 

The reason of the more chaotization of the residuals of LAGEOS vs LAGEOS 2 which had remained on orbit for shorter time period, thus, can be due to the different effect thermal thrust. For the LAGEOS satellites: $\varepsilon \, \cong \, 0.4$, $r_S \, = \, 30$ cm, $m_S \, = \, 4.1 \, \times \, 10^5$ g.  For $T \, \cong \, 280\) K and \(\Delta \, T \, = \, 5$ K the acceleration due to the thermal gradient yields  $a \, \sim \,2 \cdot  10^{-9}$ cm/sec$^{2}$ \cite{EPL}.

\begin{figure}[pb]
\centerline{\psfig{file=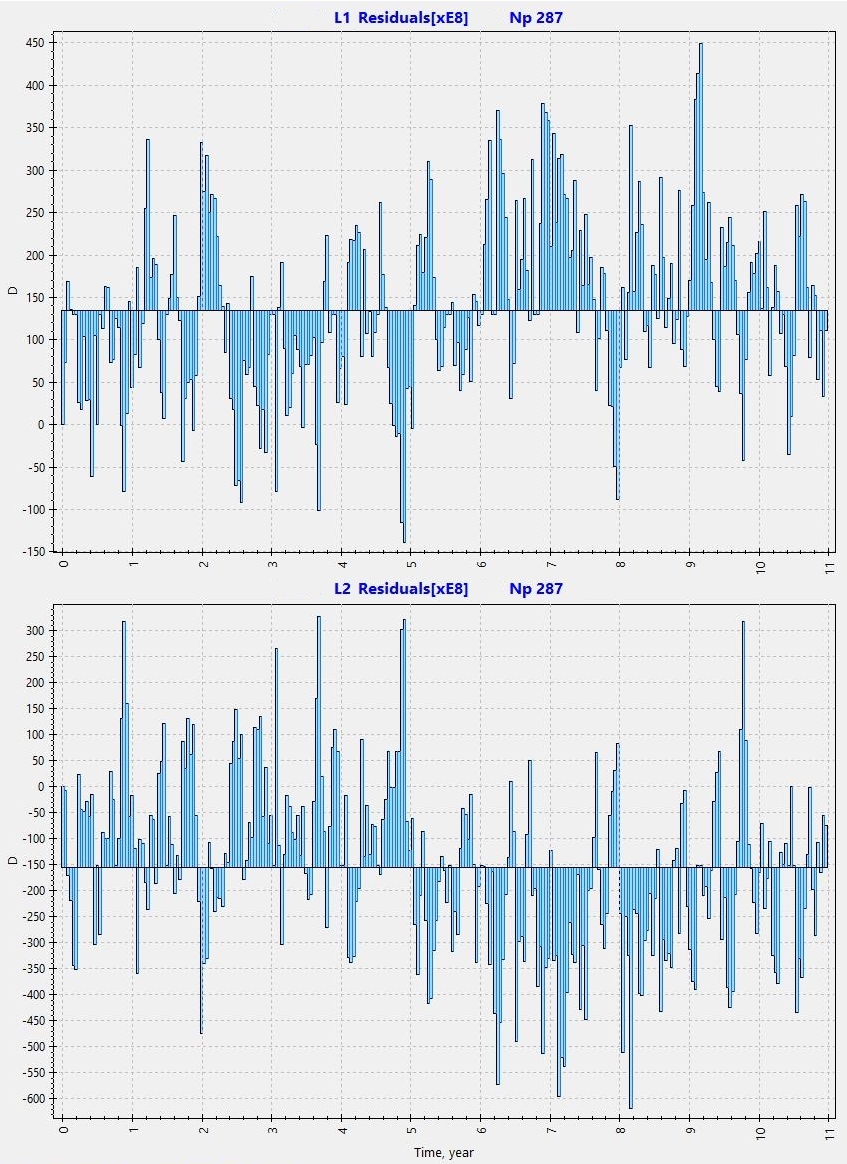,width=6.7cm}}
\vspace*{8pt}
\caption{The residuals for 11-year data of LAGEOS (upper plot) and LAGEOS-2 (bottom) satellites.}
\end{figure}

\begin{figure}[pb]
\centerline{\psfig{file=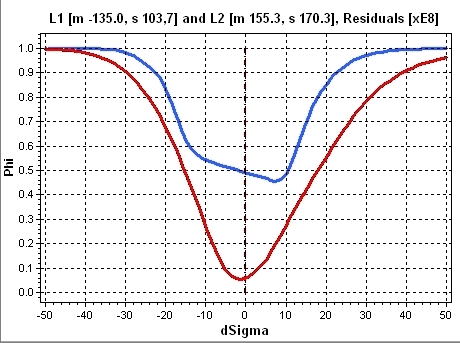,width=9.7cm}}
\vspace*{8pt}
\caption{The Kolmogorov function for the standard deviation
of the residuals of LAGEOS (upper, blue curve) and LAGEOS 2 (red).}
\end{figure}

\section{Tidal modes}

The technique for the study of Earth's gravity field perturbed by the tidal interaction of the Moon and Sun is rather developed, including e.g. the
Doodson's classification of the modes \cite{D}, in view of the practical role of the tidal effects for navigation and associated applications.      
The amplitude of the nodal shift of the orbit due to the tide modes is \cite{Kaula} 
\begin{eqnarray}
\Delta \Omega  &=&\frac{180}{\pi {\sin (i)}}\sum_{l=0}^{\infty }\sum_{m=0}^{l}%
\sqrt{\frac{GM_{\oplus }R_{\oplus }^{2}}{r^{7}(1-e^{2})}\frac{2l+1}{4\pi }%
\frac{(l-m)!}{(l+m)!}}\times   \nonumber \\
&&\times \sum_{p=0}^{l}\sum_{q=-\infty }^{\infty }\frac{dF_{lmp}(i)}{di}%
G_{lpq}(e)\frac{u_{lm}k_{lm}(\nu )}{\dot{\Gamma}_{j,m}}\sin(\nu {t}+ A_{lmpq}+\eta_{lm}(\nu)), 
\end{eqnarray}
where $M_{\oplus}$, $R_{\oplus}$ are Earth's mass and radius, respectively, $u_{lm}$ are the coefficients of the tidal
potential, $A_{lmpq}$ and $\eta_{lm}(\nu)$ are the phases independent and depending on the mode $\nu$,  $k_{lm}(\nu)$ are the Love numbers depending on the mode $\nu$ \cite{PL2010}, $F_{lmp}(i)$ is the inclination function of the perturbation and for relevant modes yields

\begin{eqnarray}
\frac{dF_{201}}{di} &=& \frac{3}{2}\sin i \, \cos i,  \nonumber \\
\frac{dF_{211}}{di} &=& -\frac{3}{2} \cos 2i,  \nonumber \\
\frac{dF_{221}}{di} &=& 3 \sin i\, \cos i,  \nonumber \\
\end{eqnarray}
the eccentricity function
\begin{equation}
G_{lpq}(e) =\frac{1}{\sqrt{1-e^2}} \simeq 1,
\end{equation}
for small eccentricity satellites, such as LAGEOS and LAGEOS 2 and, moreover, LARES.

For the parameters of LARES, i.e. orbital radius $r_L=7820\,km$, eccentricity $e_{L} = 0.0008$
period $P_{L} = 115$\, min, period of ascending node $P_{\Omega}=-211$\, days, the amplitudes of the relevant tidal modes have been computed as represented in Table 1.

\begin{longtable}{cccccc}
 Name & Mode & Love number & Period(days) & $U_{lm}$ & $\Delta\Omega (mas)$ \\\hline
 \text{    } & 055.565 & 0.315416 & 6798.3636 & 0.02793 & 5359.6967 \\
\text{    } & 055.575 & 0.313178 & 3399.1818 & -0.00027 & -25.7223 \\
 $S_a$\text{  } & 056.554 & 0.307390 & 365.2596 & -0.00492 & -49.4353 \\
 $S_{sa}$  & 057.555 & 0.305946 & 182.6211 & -0.031 & -155.0024 \\
 \text{    } & 057.565 & 0.305896 & 177.8438 & 0.00077 & 3.7487 \\
 \text{    } & 058.554 & 0.305174 & 121.7493 & -0.00181 & -6.0183 \\
 $O_1$\text{  } & 145.555 & 0.297473 & -12.8301  & -0.262210 & 84.1201 \\
 $\pi_1$\text{  } & 162.556 & 0.289961 & -77.2026  & -0.007140 & 13.4351 \\
 $P_1$\text{  } & 163.555 & 0.286921 & -97.8938  & -0.122030 & 288.1082 \\
 \text{    } & 165.545 & 0.259851 & -204.6484 & -0.007300 & 32.6308 \\
 $K_1$\text{  } & 165.555 & 0.257463 & -211.0000 & 0.368780  & -1683.9767 \\
 $\psi_1$\text{  } & 166.554 & 0.526244 & -499.6108 & 0.002930  & -64.7528 \\
 $M_2$\text{  } & 255.555 & 0.301063 & -12.0947  & 0.631920  & -84.0658 \\
 $T_2$\text{  } & 272.556 & 0.301063 & -56.5219  & 0.017200  & -10.6932 \\
 $S_2$\text{  } & 273.555 & 0.301063 & -66.8695  & 0.294000  & -216.2409 \\
 $K_2$\text{  } & 275.555 & 0.301063 & -105.5000 & 0.079960  & -92.7870 \\\hline
\caption{Amplitudes $\Delta\Omega$ and periods of perturbation of LARES satellite generated by Moon and Sun tides of the Earth.}
\end{longtable}

The mode period is 
\begin{equation}
\dot{\Gamma}_{j,m}=\frac{1}{\sum_{i=1}^{6}\frac{j_i}{\gamma_i}+m(\frac{1}{P_{\Omega}}-\frac{1}{\gamma_1})}.
\end{equation}

\section{LARES-2016: frame-dragging 5\% accuracy test}

The detection accuracy for the dragging of inertial frames with two LAGEOS satellites was 10 \% to the value predicted by GR \cite{CP,C}.  Upon collection of the laser ranging data of LARES (Fig. 3), the limit reached by the LAGEOS satellites was expected to be overpassed. 

\begin{figure}[pb]
\centerline{\psfig{file=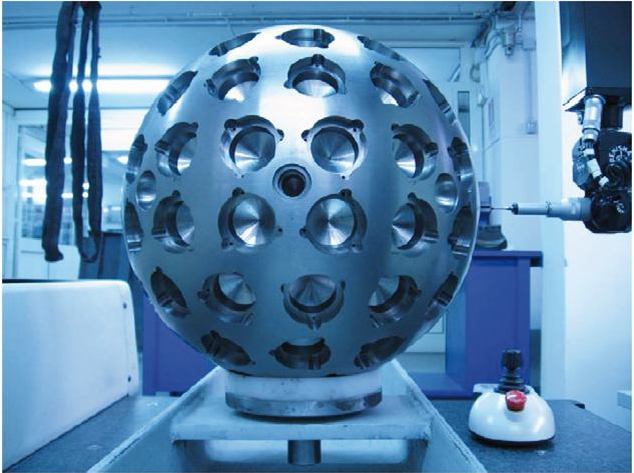,width=8.7cm}}
\vspace*{8pt}
\caption{The LARES satellite during dimensional test at OMPM (Italy) (photo from \cite{C2011}, courtesy of the Italian Space Agency).}
\end{figure}

Below, we represent the result based on the combined data of LARES, LAGEOS and LAGEOS 2 satellites collected by over 30 laser ranging stations situated worldwide during 26 February 2012 until 6 September 2015, and using GGM05S, Earth's gravity field model, produced in 2013 by means of the GRACE geodesy satellite's almost 10 years data. \cite{L2016}

Referring for details to the analysis described in \cite{L2016} and to the references therein, we mention that the GR's value for
frame-dragging effect was confirmed at the accuracy
\begin{equation}
\mu = 0.994 \pm 0.002 \pm 0.05.
\end{equation}   
Here, $\mu=1.00$ corresponds the GR's value, 0.002 is the statistical error, and the larger error, 0.05, corresponds to the systematic one determined by the uncertainty in the adopted Earth's gravity field model GGM05S.  

\section{Conclusions}

Satellites LAGEOS and LAGEOS 2 and nowdays LARES proved the profound efficiency in testing the frame-dragging effect \cite{CP,C,L2016}, by now showing no departures from General Relativity in ever increasing accuracy. The extraction of the pure effect of frame-dragging from the residuals, i.e. in the differences in the satellite's measured orbits and the theoretical ones, obviously, needs refined analysis of various perturbations.
We used the Kolmogorov stochasticity parameter technique to reveal the chaotic component in the signal of LAGEOS which was longer in the orbit than the otherwise identical LAGEOS 2 and hence was differently affected by the thermal thrust (Yarkovsky effect). Thus, the study of the randomness properties of the residuals can lead to revealing of such tiny contributions in the signals. 

We computed the contribution of the dominant set of tidal modes of the Earth's gravity on the orbit of LARES. The represented set of modes ensures the obtained 5\% accuracy for the frame-dragging \cite{L2016}. 

More refined analysis of the non-gravitational effects and, especially, the possibility for future satellite missions with better characteristics will be crucial in tracing the actual limits of General Relativity and of the linked fundamental physical principles.

\end{document}